\documentstyle[prl,preprint,aps]{revtex}

\begin{document} 
\tighten
\draft 
\preprint{ } 
\title{Pairing and isospin symmetry in proton-rich nuclei} 
\author{J. Engel$^1$, K. Langanke$^2$ and P. Vogel$^3$}  
\address{$^1$Department of Physics and Astronomy, CB3255,
University of North Carolina, \\Chapel Hill, North Carolina 27599 \\
$^2$ W. K. Kellogg Radiation Laboratory, 106-38, California
Institute of Technology\\ Pasadena, California 91125 \\
$^3$ Physics Department, California
Institute of Technology, Pasadena, California 91125  
} 
\date{\today} 
\maketitle

\begin{abstract} 

Unlike their lighter counterparts, most odd-odd $N=Z$ nuclei with mass $A >
40$ have ground states with isospin $T$=1, suggesting an increased role for
the isovector pairing interaction.  A simple SO(5) seniority-like model of
this interaction reveals a striking and heretofore unnoticed interplay between
like-particle and neutron-proton isovector pairing near $N=Z$ that is
reflected in the number of each kind of pair as a function of $A$ and $T$.
Large scale shell-model calculations exhibit the same trends, despite the
simultaneous presence of isoscalar pairs, deformation, and other correlations.
\end{abstract} 


With the advent of radioactive beams, unstable nuclei on the proton-rich side
of the valley of stability are receiving increased attention.  These nuclei
will surely exhibit larger neutron-proton ($np$) pairing effects than do heavy
stable nuclei, in which the valence protons and neutrons lie in different
shells.  This observation has been difficult to precisely quantify,
however\cite{r:Camiz,r:Faessler}, despite many years of $np$-pairing
theory\cite{r:Goodman}.  The reason, as we shall see, is that in heavy nuclei
with $N$ near $Z$, the delicate balance between like-particle ($nn$ and $pp$)
and $np$ pairing has thus far partly eluded standard approximations.

In this work we focus our attention on isovector pairing in the ground states 
of $fp$-shell nuclei with $N$ not too much bigger than $Z$.  Low-lying excited 
states are also interesting, but ground states deserve special attention 
because they determine masses and lifetimes. Our primary motivation for 
addressing isovector pairing is that unlike their lighter counterparts odd-odd 
$fp$-shell nuclei with $N=Z$ usually have ground-state isosopin $T=1$.  This 
suggests that the ground-state effects of isovector pairing are larger in the 
heavier nuclei, and perhaps not only when $N$ is precisely $Z$.  Although 
isoscalar pairing obviously should not be neglected where $np$ correlations 
are likely to be important, isovector pairing, which also generates $np$ 
correlations, clearly deserves close examination in proton-rich nuclei.

Isospin symmetry dicates the relative strengths of the $nn$, $pp$, and $np$
parts of the isovector pairing interaction.  Unfortunately the generalized
Hartree-Fock-Bogoliubov (HFB) methods typically used to treat pairing break
isospin symmetry, and except in the $N=Z$ nuclei the lowest-energy $T=1$ pairs
usually contain no $np$ correlations\footnote{The only exceptions of which we
are aware are in Ref.\ \cite{r:Faessler}, but there the authors break isospin
symmetry completely by using $T=1$ $np$ pairing to represent $T=0$ pairing.}.
Even in $N=Z$ nuclei the method tends not to mix the $np$ isovector pairs with
the like-particle pairs, giving two degenerate states instead.  Although we
argue later that projecting HFB wave functions onto states with good isospin
will allow such mixing, this procedure has never been systematically
implemented\cite{r:Weneser} and the role played by $T=1$ $np$ pairing is
therefore not fully appreciated.  Here we examine the interplay between the
different kinds of isovector pairs with techniques that conserve isospin from
the start.  We will see that like-particle and $np$ pairing compete in a
subtle but clearly visible fashion.

We proceed essentially in two steps.  First we analyze the situation in an
exactly solvable model that gives us some insight into precisely how the
different kinds of pairs compete.  We then show that the same patterns of
competetion are also present in realistic large-basis shell model
calculations.  The similarity between the results of a simple model and those
of complicated numerical calculations that include many other effects besides
isovector pairing strongly suggests that the patterns we describe here are
quite general, and should also be present in real nuclei.  After discussing 
how HFB theory might be altered slightly to capture these patterns, we 
conclude with a brief discussion of possible evidence for simple 
$np$ isovector-pairing effects in binding energies.

We begin with the simple isovector-pairing model.  We 
consider a single $j$ orbital or a degenerate set with a total of $2 \Omega$ 
$m$-substates and the pure isovector pairing interaction
\begin{eqnarray} 
\label{e:H} 
H &=& -G \ \vec{S}^{\dag} \cdot \vec{S}  ~, \\ 
{S}^{\dag}_{M} & = & \sum_{j,m > 0} \ [ a^{\dag}_m
a^{\dag}_{\bar{m}} ]^{T=1}_M ~,  \nonumber 
\end{eqnarray} 
where $m$ is the $z$-projection of the angular momentum in a single-particle
state, $\bar{m}$ labels the corresponding time-reversed state, $G$ is an
arbitrary constant, and the square brackets indicate isospin coupling, so
that ${S}^{\dag}_{M}$ is an isovector pair creation operator with projection
$M$.  Together with the isospin and number operators, the
angular-momentum-zero pair creation and annihilation operators above form the
algebra $SO(5)$.  The Hamiltonian in Eq.\ (\ref{e:H}), a generalization of
the like-particle SU(2) seniority interaction, is solvable analytically.  The
space of fully-paired states in which the Hamiltonian acts is spanned by the
set $|\Omega, {\cal N},T,T_z\rangle$, where ${\cal N} = n/2$ is half the
total number of particles, $T$ is the nuclear isospin, and $T_z = 1/2(N-Z)$
is its z-projection.  Algebraic techniques for computing matrix elements of
operators between these states have existed for some
time\cite{r:Hecht1,r:Ginn,r:Hecht2}.

To quantify the effects of the different pairing modes, we define the 
operators
\begin{equation}
\label{e:pairs} 
{\cal N}_{pp} = \frac{S^{\dag}_1 S_1}{\Omega}~, \ \ \ {\cal N}_{nn} = 
\frac{S^{\dag}_{-1} S_{-1}}{\Omega}~, \ \ \ {\cal N}_{np} = \frac{ S^{\dag}_0 
S_0~}{\Omega},
\end{equation}
the sum of which enters $H$ in Eq.\ (\ref{e:H}).  These operators are rough
measures of the numbers of $nn$, $pp$ and $np$ pairs; 
their expectation values are related to the usual pairing gaps of the HFB
theory.  In the limit of large $\Omega$, the pair-creation operators 
$S^{\dag}_j$ and their conjugates obey boson
commutation relations, and the three operators just defined count the number
of different kinds of bosons.  To leading order in $1/\Omega$, the
expectation values of these ``number operators" therefore sum to ${\cal N}$
and reproduce the results of Ref.\ \cite{r:Halse}, in which similar
quantities (e.g.\ the total number of $np$ bosons) are computed in the
$SU(3)$ limit of the $IBM$-4.  With the generalized Wigner-Eckart theorem and
the $SO(5)$ Wigner coefficients tabulated in \cite{r:Hecht2}, we can obtain
exact matrix elements for these operators, valid to all order in $1/\Omega$,
between states with arbitrary values of ${\cal N}, T$, and $T_z$.  All
even-even nuclei, which we examine first, have ground-state isospins $T =
T_z$; for these the ``number operators" have the ground state expectation
values (with $N \ge Z$)
\begin{eqnarray}
\label{e:expect}
\langle {\cal N}_{np} \rangle &= &\frac{\left({\cal N}-T \right) 
\left(1-\frac{{\cal 
N}-T-3}{2 \Omega} \right)}{2T+3} ~, \nonumber \\
\langle {\cal N}_{pp} \rangle& = &(T+1) \langle {\cal N}_{np} \rangle ~, \\
\langle {\cal N}_{nn} 
\rangle& =& \langle {\cal N}_{pp} \rangle + T(1-\frac{{\cal N}-1}{\Omega}) ~. 
\nonumber
\end{eqnarray}

These expressions are striking.  Not surprisingly, when $T=0$, i.e.\ $N=Z$,
all three expectation values are equal.  But adding just a single pair of
neutrons so that $T=T_z=1$ causes the number of $np$ pairs to drop by about
40\%, while the number of $pp$ pairs actually increases by about 7\% (the
number of $nn$ pairs, naturally, increases even more).  This result, which in
the boson limit is due solely to the algebra of the
3-dimensional harmonic oscillator, fully captures the subtle competition
between the three pair condensates.  Increasing the number of
neutron pairs increases the collectivity of the neutron condensate, making
fewer neutrons available to pair with protons.  As a result the protons pair
more often with one another, even though their number hasn't increased, and
the binding in the $np$ condensate drops dramatically.  The right side of
Fig.\ 1 shows what happens as $T$ gets larger still:  the trends continue to
very large $T$, where the state is nearly a product of $nn$ and $pp$
condensates.  These results contrast with those of 
existing calculations\cite{r:Camiz,r:Goodman}, which as we have noted rarely 
exhibit
coexistence between like-particle and $np$ isovector pairs (Ref.\ 
 \cite{r:Camiz} applies HFB theory to the same model describe here).

Interestingly, a generalization of $SO(5)$ seniority to an $SO(8)$-based
model that includes isoscalar pairing\cite{r:Pang} shows the pattern of
isovector pairs to be relatively insensitive to the strength of the new mode,
unless the latter is completely dominant\cite{r:EPSV}.  In other words, the
relative numbers of isovector $np$, $pp$, and $nn$ pairs don't change
substantially (though the absolute numbers can) from the predictions above
when an entirely different interaction is added, suggesting that the
footprints of isovector pairing are not easy to erase.  Realistic shell-model
calculations in proton-rich $fp$ shell nuclei bear this prediction out
remarkably.  Such calculations are difficult because of the size of the $fp$
model space\cite{r:Zuker}, and only with the advent of the Shell Model Monte
Carlo (SMMC) method have they begun to include the entire shell for nuclei
heavier than $A=50$\cite{r:Koonin}.  Performed with the modified Kuo-Brown
interaction KB3, the SMMC calculations reproduce the general properties of
these nuclei quite well\cite{r:Langanke1}.  The calculated
total $B(E2)$ and Gamow-Teller strengths agree particularly well with data,
indicating that both isovector and isoscalar pairing are realistically
described.  Alongside the scaled predictions of the simple SO(5) model, we
display in Fig.\ \ref{f:1} results of the SMMC (with the same input and
computing procedure as in Refs.\ \cite{r:Langanke1,r:Langanke2}) for the
``number" of $nn$, $pp$, and $np$ isovector pairs multiplied by $\Omega$ in
the Cr and Fe isotopes.  Although the curves produced by the two models are
not identical, they are strikingly similar; the additional physics in the
shell model calculation --- several orbits, isoscalar pairing, spin-orbit
splitting, long-range correlations, deformation, etc.\ --- reduce the total
number of isovector pairs without dramatically changing their relative
number.  [The kinks in the SMMC ${\cal N}_{nn}$ curves reflect the closing in
$^{54}$Fe and $^{52}$Cr of the $f_{7/2}$ subshell.]  The stubborn persistence
of the patterns in all kinds of calculations strongly suggests that real
nuclei exhibit them as well.

What about the $N=Z$ nuclei in the $fp$ shell, the ground state isospins of
which are the clearest signal of $np$ isovector pairing?  How is the fact that 
the odd-odd nuclei have $T=1$ reflected in the $nn$, $pp$, and $np$ 
condensates?  Intuition suggests that the extra
isovector neutron-proton pair should give the $np$ mode a slight advantage in
its competition with like-particle pairing, and in the seniority model it
clearly does.  For states with $T=T_z+1$, the model predicts
\begin{eqnarray}
\label{e:expect1}
\langle {\cal N}_{np} \rangle & = & \frac{\left(3{\cal N}+3-T 
\right) \left(1-\frac{{\cal 
N}-T-3}{2 \Omega}\right)}{2T+3}-\frac{{\cal 
N}+T+1}{2 \Omega}~, \nonumber \\ 
\langle {\cal N}_{pp} \rangle & = & \frac{\left({\cal N}-T 
\right) T \left(1-\frac{{\cal N}-T-3}{2 \Omega}\right)}{2T+3}  ~,  \\ 
\langle {\cal N}_{nn} \rangle & = & \frac{\left[ \left({\cal 
N}+T+1 \right)T-3 \right] \left(1-\frac{{\cal 
N}-T-3}{2 \Omega} \right)}{2T+3} - \frac{(T-1)({\cal N}+T+1)}{2 \Omega} ~. 
\nonumber 
\end{eqnarray}
In the odd-odd $N=Z$ nuclei with $T=1$, these expressions imply more $np$
pairing than in the even-even $N=Z$ nuclei.  Fig.\ \ref{f:2} shows $\langle
{\cal N}_{np} \rangle$ and 
$\langle {\cal N}_{nn} \rangle$ = $\langle {\cal N}_{pp} \rangle$ 
as a function of
mass number when $N=Z$; a characteristic odd-even staggering from the
additional $np$ pair in the odd-odd nuclei is the salient feature.  As
before, we find that this pattern also appears, though less prominently, in
the SMMC calculations, the results of which are displayed in the same figure.
Isoscalar pairing, which by itself produces smoother curves, may be partly 
responsible for deviations from the simple formulae in Eq.\ 
(\ref{e:expect1}), but it does not erase the staggering produced by isovector 
pairing.  

As noted above, the reason isovector $np$ correlations have gone largely
unnoticed is that HFB theory as applied thus far does not mix them with $nn$
and $pp$ pairs.  It also rarely lets isoscalar and isovector pairs
coexist\cite{r:EPSV,r:Sauer}, instead (usually) assigning each $N \approx Z$
nucleus three separate states:  one with no $np$ pairs, the second composed
entirely of $T=1$ $np$ pairs and degenerate with the first, and the third 
containing only $T=0$
pairs\cite{r:EPSV,r:Sauer} and lying higher or (usually) lower in energy.  
Since the characteristic odd-even staggering results from mixing the first two 
solutions, it is not present in any of the three individually.  We strongly 
suspect, however, that the restoration of isospin symmetry in HFB theory will 
at least partly correct these problems.  To verify this conjecture we have 
reexamined the HFB treatment of SO(5), the existing version of which shows no 
neutron-proton correlations\cite{r:Camiz}.

Restoring the symmetry means viewing the symmetry-violating quasiparticle
vacuum as an intrinsic state in isospin space.  From this vantage point $T_z$
plays the role of the rotational quantum number $K$ that labels bandheads.
The usual constraint that $<T_z> = N-Z$ (the analog of the rotational quantum
number $M$) should therefore be eliminated or replaced.  Here we assume
``axial symmetry", i.e.  an intrinsic HFB wave function of the form
\begin{equation}
| {\rm HFB (intr)}~ \rangle = {\rm exp}~[\alpha S^{\dag}_0] ~|0\rangle~,
\end{equation}
with $\alpha^2$ fixed at $2{\cal N}/(2\Omega-\cal{N})$ by the usual 
constraint on the average total nucleon number. We project approximately by 
multiplying the intrinsic state by a ``collective" rotational wave function 
$D^T_{T_z,0}(\Omega)^*$ (where $\Omega$ now stands for a rotation), in 
analogy with the Nilsson model.  Continuing the analogy we assign the 
``number operators" intrinsic and collective parts as well, and for $T=T_z$ 
obtain, e.g., 
\begin{equation}
\label{e:expectHFB}
{\rm projected~HFB:} ~~\langle {\cal N}_{np} \rangle  \approx \frac{{\cal N} 
\left(1-\frac{{\cal N}}{2 \Omega} \right)}{2T+3} + {\cal O}({\cal 
N}^2/\Omega^2) 
 ~.
\end{equation}
This result makes perfect sense because the Nilsson prescription approximates
projection only when ${\cal N} , \Omega >> T$.  The last term above, which
also appears in the BCS approximation for like-particle pairing, is therefore
much smaller than the others.  Complete projection should add the missing
factors of $T$ so that the exact expression for ${\cal N}$ in Eq.\
(\ref{e:expect}) is reproduced up to terms of order 1 or less.  This kind of
procedure may also facilitate the more complicated dynamical mixing of $T=1$
and $T=0$ pairs\cite{r:Wyss}, in part because eliminating the traditional
constraint on $T_z$ allows a more general quasiparticle vacuum.  Isospin
projection in HFB theory clearly merits further study.

In concluding we return to the physical results obtained above:  $fp$-shell
nuclei exhibit striking and easily understood variations in the numbers of
isovector $nn$, $pp$, and $np$ pairs.  We strongly suspect that the
competition between the different kinds of pairs affects measurable
properties of real nuclei.  Unfortunately the heavy proton-rich nuclei are
quite short-lived and few data are available to confirm this belief.  Some
structural information, however, can be gleaned solely from binding energies,
which beside lifetimes are the most systematically measured quantities in
these unstable nuclei and should be affected by pairing.  A recent
paper\cite{r:Piet} points out that in $sd$-shell and $p$-shell nuclei the
quantity $\delta V_{np} (N,Z)$, defined as
\begin{equation} 
\delta V_{np} (N,Z) \equiv \frac{1}{4} \left\{[B(N,Z) - B(N-2,Z)] - [B(N,Z-2) 
- B(N-2,Z-2)]\right\} ~, 
\label{eq:su4} 
\end{equation}
where $B$ means binding energy and $N,Z$ are both even (similar quantities
can be constructed for the other nuclei) is much larger when $N = Z$ than
when $N \ne Z$.  The paper then shows that this pattern follows from $SU(4)$
symmetry and argues that though apparently broken badly, primarily by
spin-orbit splitting, the symmetry is therefore in some sense good.  Here we
note first that the combination in Eq.\ (\ref{eq:su4}) is also enhanced for
$N=Z$ in the heavier nuclei with $40 \le A \le 64$, so that the argument's
range of validity is even larger than claimed.  We find in addition, however,
that the preference for $N = Z$ is not an exclusive feature of SU(4)
symmetry; the same nuclei are in fact equally preferred in the pure
isovector-seniority model discussed above.  Though this fact doesn't by
itself prove that isovector neutron-proton pairing plays the visible role
presented here, it is nonetheless suggestive and should motivate a greater
effort to understand the properties of proton-rich nuclei, especially in
regions that are still outside the ambit of shell-model calculations.  The
nuclei at the edge of stability will come under increasing experimental
scrutiny, and the apparent validity of simple models (and dynamical
symmetries) makes them all the more interesting.

We thank S.\ Pittel and P.\ van Isacker for informative discussions about HFB
theory and $SU(4)$, and the SMMC collaboration for allowing us the use of
some of their results prior to publication.  We were supported in part by the
U.S.\ Department of Energy under grants DE-FG05-94ER40827 and
DE-FG03-88ER-40397, and by the U.S.\ National Science Foundation under grants
PHY94-12818 and PHY94-20470.

\begin{figure}
\caption[Figure 1]{The quantities $\Omega \langle {\cal N}_{pp} \rangle$,
$\Omega \langle {\cal N}_{nn} \rangle$, and $\Omega \langle {\cal N}_{np} 
\rangle$ for Fe and Cr isotopes, as a function of $N-Z = 2T_z$. On the left 
are the results of the Shell Model Monte Carlo calculation, and on the right 
are those from Eq. (\ref{e:expect}) (with $\Omega = 10$, half the number of 
single-particle levels in the $fp$ shell) scaled by a factor 0.5 to account 
for the reduction of the single-particle level density near the Fermi level 
and other effects.}
\label{f:1}
\end{figure}

\begin{figure}
\caption[Figure 2]{The quantities $\Omega \langle  {\cal N}_{pp} \rangle 
\equiv \Omega \langle  {\cal N}_{nn} \rangle$ (top panel), and $\Omega 
\langle {\cal N}_{np} \rangle$ (bottom panel) for $N=Z$ nuclei. The full line 
connecting the points with error bars is the result of the Shell Model Monte 
Carlo calculation, and the dashed line is from Eqs. (\ref{e:expect}) and 
(\ref{e:expect1}), again scaled by a factor 0.5.}
\label{f:2}
\end{figure}

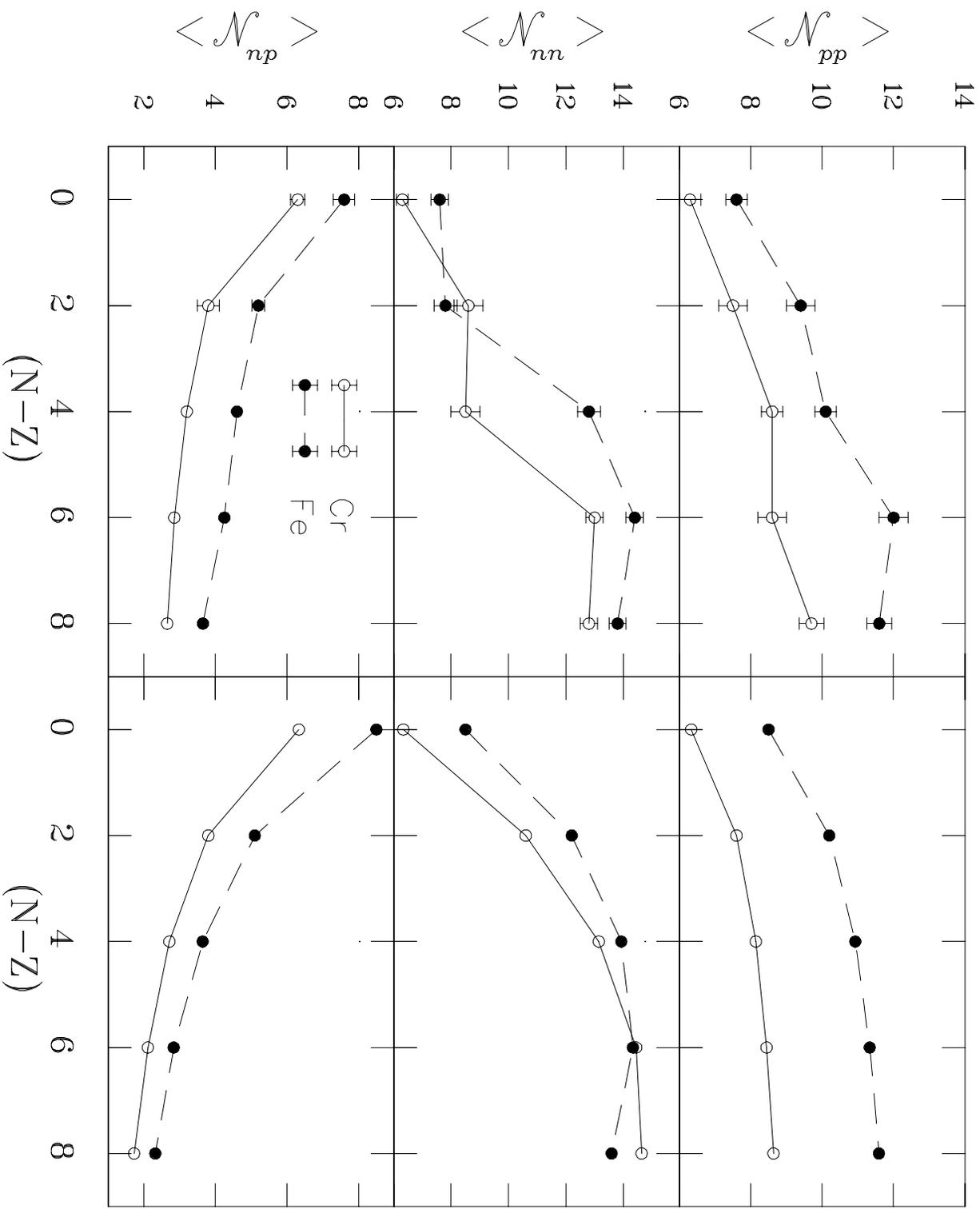
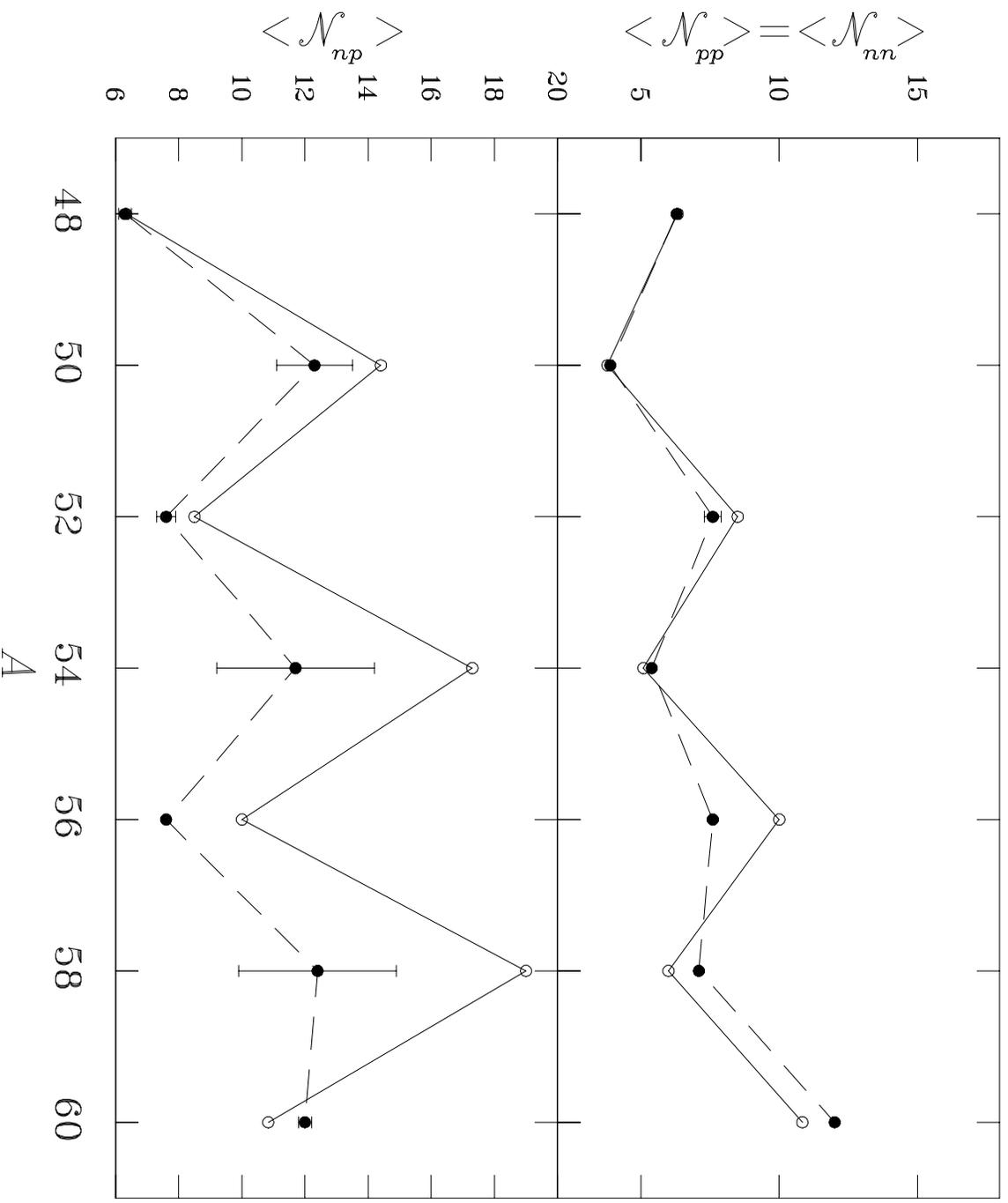
\end{document}